\documentclass[english,aps,prd,a4paper,groupedaddress,preprintnumbers,floatfix,nofootinbib,twocolumn]{revtex4-1}
\usepackage{graphicx}
\usepackage{url}
\usepackage{xcolor}
\usepackage{hyperref}
\hypersetup{colorlinks=true,linkcolor=redLinks,citecolor=greenLinks,
urlcolor=redLinks,
pdfborder={0 0 1}}
\hypersetup{
    bookmarks=true,         
    unicode=false,          
    pdftoolbar=true,        
    pdfmenubar=true,        
    pdffitwindow=false,     
    pdfstartview={FitH},    
    pdftitle={My title},    
    pdfauthor={Author},     
    pdfsubject={Subject},   
    pdfcreator={Creator},   
    pdfproducer={Producer}, 
    pdfkeywords={keyword1, key2, key3}, 
    pdfnewwindow=true,      
    colorlinks=false,       
    linkcolor=red,          
    citecolor=green,        
    filecolor=magenta,      
    urlcolor=cyan           
}
\usepackage{amsmath}
\usepackage{amssymb}
\colorlet{shadecolor}{gray!15}
\definecolor{greenLinks}{rgb}{0, 0.6, 0} 
\definecolor{blueLinks}{rgb}{0, 0, 0.6}
\definecolor{redLinks}{rgb}{0.6, 0, 0}
\definecolor{tempText}{rgb}{0.55, 0.10,0.67}
\definecolor{eprintLinks}{rgb}{0.4, 0.4, 0.4}
\definecolor{journalLinks}{rgb}{0.6, 0, 0}
\newcommand{\MYhref}[3][redLinks]{\href{#2}{\color{#1}{#3}}}%

\usepackage[T1]{fontenc} 

\usepackage{hyperref}

\newcommand {\black} {\color{black}}

\def\21{$\mathrm{SU(2)_L \otimes U(1)_Y}$ }
\def\lfv{lepton flavour violation }
\def\lfc{lepton flavour conservation }
\def\lnv{lepton number violating }

\newcommand{\Cinvestav}{Departamento de F\'{\i}sica, Centro de
  Investigaci{\'o}n y de Estudios Avanzados del IPN\\ Apdo. Postal
  14-740 07000 Mexico, DF, Mexico}

\newcommand{\AddrAHEP}{AHEP Group, Institut de F\'{i}sica Corpuscular --
  C.S.I.C./Universitat de Val\`{e}ncia, Parc Cientific de Paterna.\\
  C/Catedratico Jos\'e Beltr\'an, 2 E-46980 Paterna (Val\`{e}ncia) - SPAIN}

\begin{document}


\title{ New ambiguity in probing CP violation in neutrino  oscillations}

\author{O. G. Miranda~$^1$}\email{omr@fis.cinvestav.mx}
\author{M. T\'ortola~$^2$}\email{mariam@ific.uv.es}
\author{J. W. F. Valle~$^2$} \email{valle@ific.uv.es, URL:
  http://astroparticles.es/} 
\affiliation{$^1$~\Cinvestav}
\affiliation{$^2$~\AddrAHEP}

\begin{abstract}

  If neutrinos get mass via the seesaw mechanism the mixing matrix
  describing neutrino oscillations can be effectively nonunitary. We
  show that in this case the neutrino appearance probabilities involve
  a new CP phase $\phi$ associated with nonunitarity.
  This leads to an ambiguity in extracting the ``standard''
  three--neutrino phase $\delta_{CP}$, which can survive even after
  neutrino and antineutrino channels are combined.
  Its existence should be taken into account in the planning of any
  oscillation experiment aiming at a robust measurement of  $\delta_{CP}$.

\end{abstract}


\maketitle
\section{Introduction}

The celebrated discovery of neutrino oscillations and the precision
measurements of the corresponding parameters have opened a new era in
particle physics.
So far experiments have measured two neutrino mass differences and
three mixing angles~\cite{Maltoni:2004ei}. Four out of these
measurements are very
precise~\cite{Forero:2014bxa,Capozzi:2016rtj,Gonzalez-Garcia:2015qrr},
while the octant of the atmospheric mixing angle $\theta_{23}$ still
remains uncertain.
In order to complete such simple three--neutrino paradigm, the hunt for
leptonic CP violation stands out as the next challenge, taken up by
experiments such as T2K and NO$\nu$A aimed at determining the Dirac CP
phase\footnote{The so--called Majorana phases~\cite{Schechter:1980gr}
  do not affect the oscillation probabilities, only \lnv
  processes~\cite{Schechter:1981gk,Doi:1980yb,Branco:2011zb}.}
$\delta_{CP}$.
It has long been noted, however~\cite{Schechter:1980gr}, that such a
simple closed picture holds true only for the simplest benchmark, in
which there are just the three families of conventional orthonormal
neutrinos.

One of the most popular ways to induce neutrino mass is the (type-I)
seesaw mechanism~\cite{Minkowski:1977sc,Yanagida:1979,
  Mohapatra:1980ia,Schechter:1980gr,Gell-Mann:1980vs,
  Valle:2015pba}. The latter invokes the tree--level exchange of
heavy, so far undetected, ``right-handed'' neutrinos.
Such messenger particles may be accessible at the Large Hadron
Collider~\cite{delAguila:2007qnc,Boucenna:2014zba,Das:2012ii,Deppisch:2015qwa}. In this case they are
expected to couple in the charged current with appreciable strength,
leading to a rectangular form of the mixing matrix characterizing the
leptonic weak interaction~\cite{Schechter:1980gr}.
The outcome is that the effective mixing matrix describing neutrino
oscillations will not in general be unitary. As a result more
parameters are required in order to fully describe neutrino
oscillations, posing an important challenge for future neutrino
experiments~\cite{Escrihuela:2015wra,Miranda:2016ptb}.

In this letter we focus on the description of neutrino oscillations
with nonunitary neutrino mixing matrix, particularly on the role of
the extra CP phase required to describe oscillations under this
hypothesis.
In order to carry out this study, we find it most convenient to make
use of the original symmetric parametrization~\cite{Schechter:1980gr}
of the neutrino mixing matrix~\cite{Rodejohann:2011vc}, in which the
possible ``confusion'' between the ``standard'' and ``new'' CP
violating phase combinations in the neutrino oscillation probability
can be clearly seen.
We illustrate this new ambiguity in extracting the Dirac CP phase for
different $L/E$ choices and different values of the new parameters
characterizing nonunitarity.
The ambiguities we find are genuinely new, without a counterpart within
the standard three--neutrino oscillation paradigm~\footnote{ They add to
  the well known ambiguities associated to the mass hierarchy and
  $\theta_{23}$
  octant~\cite{Fogli:1996pv,Minakata:2001qm,Barger:2001yr,Ghosh:2015ena}.}.

We would like to stress that the extra CP phase leading to the
one--parameter degeneracy in the neutrino conversion rates constitutes
a natural feature of neutrino oscillations within a broad class of
seesaw theories~\cite{Miranda:2016ptb}.
The effects of these new degeneracies will have to be taken into
account in the planning of current and upcoming experiments aiming at
a robust determination of the leptonic Dirac CP violation phase
$\delta_{CP}$, such as T2K, NO$\nu$A, DUNE, MOMENT, etc.

\section{new degeneracies in oscillations} 
In the presence of heavy neutral leptons, the mixing matrix describing
the leptonic weak interactions will be a rectangular $3\times(3+m)$ 
matrix~\cite{Schechter:1980gr}, $K$, with $m$ denoting the number of
heavy states.
As a result, the effective $3\times3$ mixing submatrix describing
neutrino oscillations will be non--unitary.
Using the original symmetric form in~\cite{Schechter:1980gr} one can
write the latter, in full generality, as~\cite{Escrihuela:2015wra}
\begin{equation} 
N=\left(\begin{array}{ccc}\alpha_{11} & 0 & 0\\
\alpha_{21} & \alpha_{22} & 0\\
\alpha_{31} & \alpha_{32} & \alpha_{33}
\end{array}\right)\: U^{3\times3}
\label{eq:Ndescopm_C1} ,
\end{equation} 
where $U^{3\times3}$ is the usual three--neutrino unitary mixing matrix.
The description of unitarity violation involves three real parameters,
$\alpha_{ii}$, that should be close to one, and three small complex
off-diagonal parameters, $\alpha_{ij}$.
Within such a nonunitary framework the neutrino appearance probability
\textit{in vacuo}, $P_{\mu e}$, will be similar in form to that found
in the unitary case, but with $U$ replaced by the matrix
$N$~\footnote{Expressions for $P_{ee}$ and $P_{\mu\mu}$ were
    given in Ref.~\cite{Escrihuela:2015wra}. For such CP conserving
    channels nonunitarity hardly affects the determination of
    oscillation parameters, which are rather robust.}.

This probability can be simplified by neglecting the cubic products of
the small parameters $\alpha_{21}$, $\sin\theta_{13}$ and
$\sin(\frac{\Delta m^2_{21}L}{4E})$.
In this case the previous expression reduces to the very simple and
compact master formula~\cite{Escrihuela:2015wra}
\begin{eqnarray} 
\label{eq:Pmue} 
P_{\mu e} = 
 \alpha_{11}^2\alpha_{22}^2 P^{3\times3}_{\mu e}
+  \alpha_{11}^2 \alpha_{22}|\alpha_{21}|  P^{I}_{\mu e} 
+ \alpha_{11}^2|\alpha_{21} |^2 , \\[-.4cm] \nonumber
\end{eqnarray}
where the new physics information related to the seesaw mechanism is
encoded in the $\alpha$--parameters describing non--unitarity, coming
from Eq~(\ref{eq:Ndescopm_C1}). Here we have used the original
symmetric parametrization of the lepton mixing
matrix~\cite{Schechter:1980gr} and denoted the standard
three--neutrino conversion probability by $P^{3\times 3}_{\mu e}$. The
latter is given explicitly in
Refs.~\cite{Freund:2001pn,Akhmedov:2004ny,Nunokawa:2007qh}.
\begin{figure}[!b]
\centering
\includegraphics[width=0.42\textwidth]{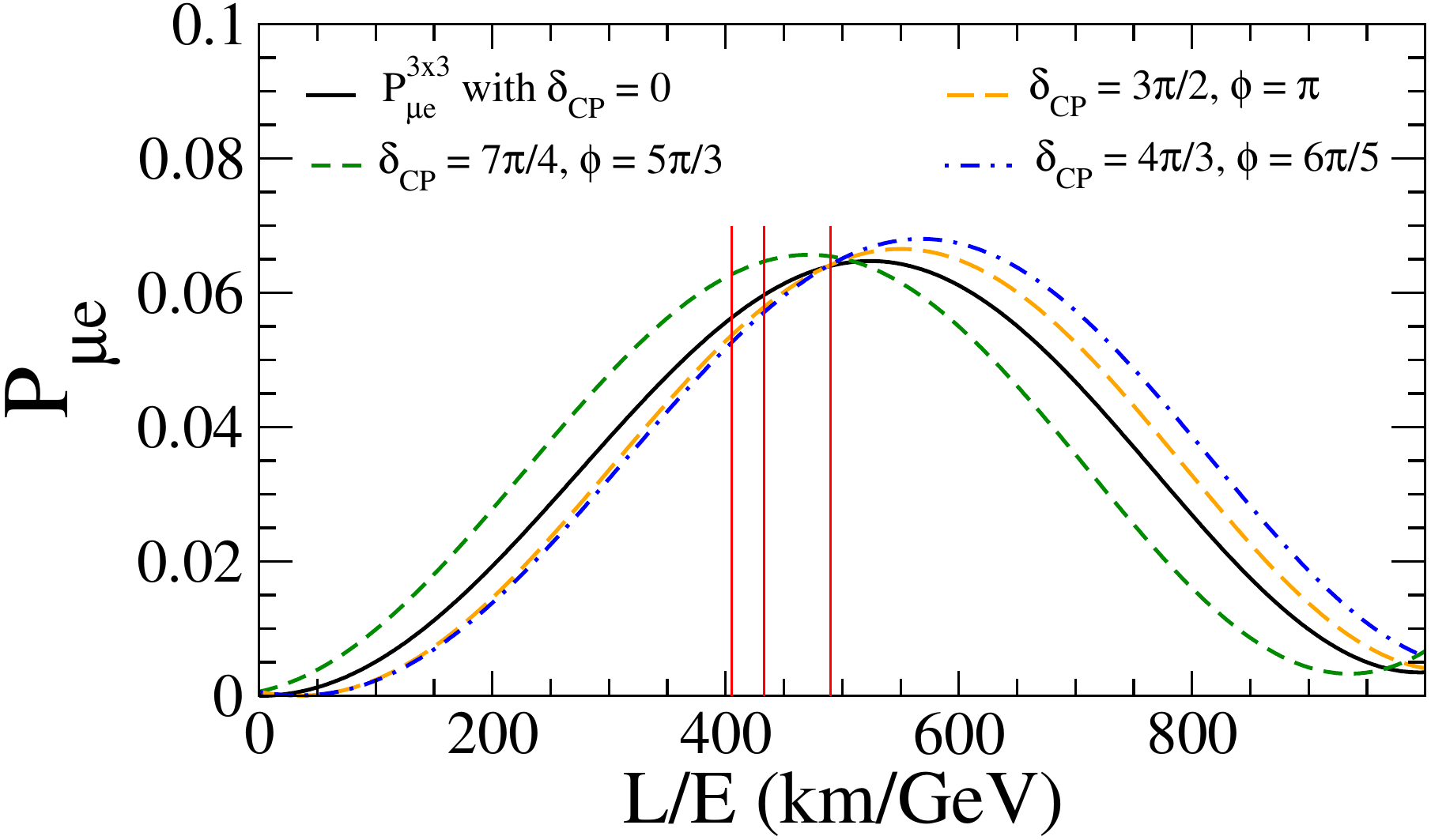}
\includegraphics[width=0.4\textwidth]{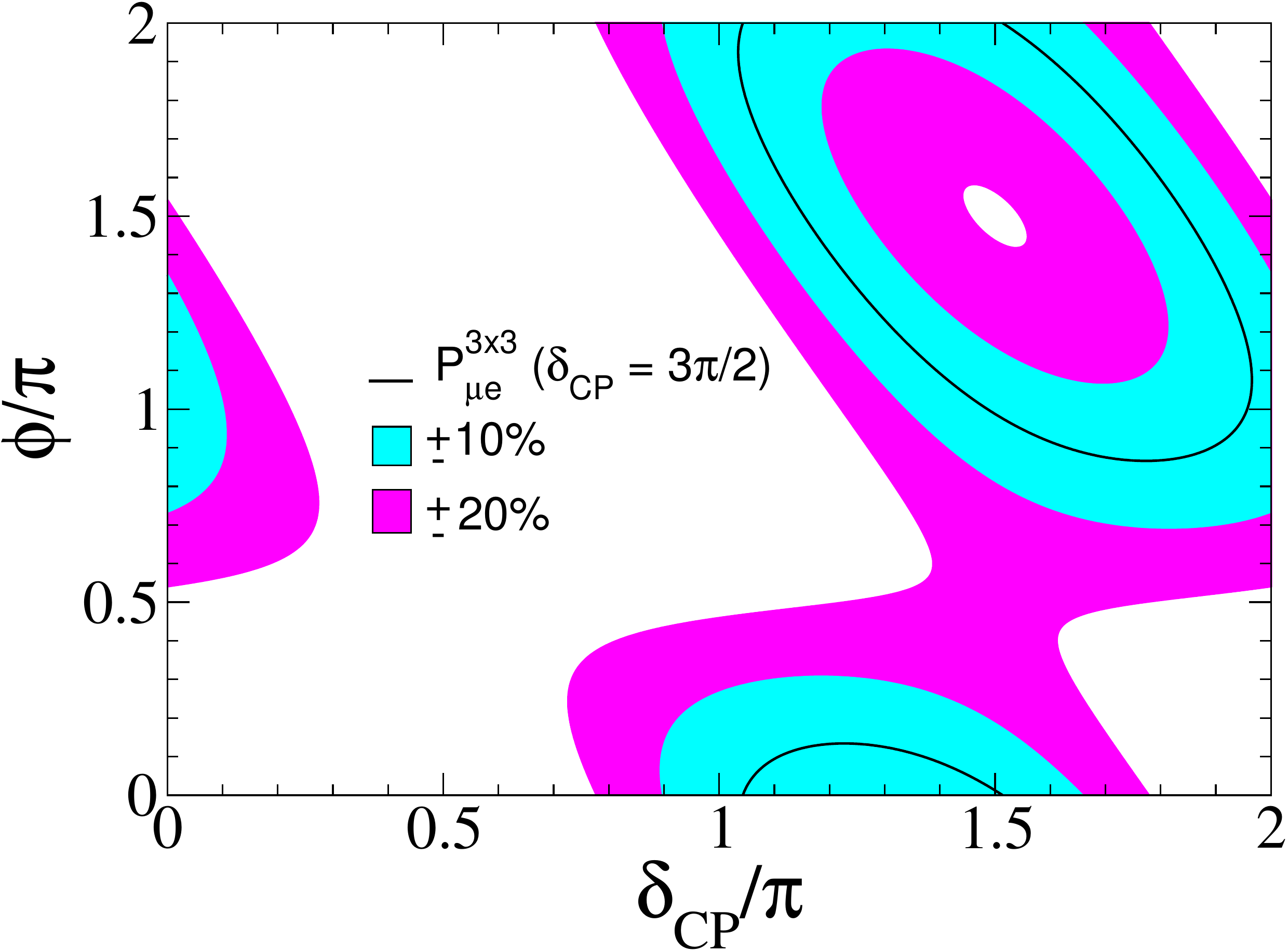}
\caption{\label{fig:pemu-deg} Top: Vacuum appearance probability
  P$_{\mu e}$ versus $L/E$ for different phase combinations,
  illustrating a degeneracy for $L/E$ = 500 km/GeV. Vertical lines indicate 
  the mean value of $L/E$  for  NO$\nu$A (405), DUNE (433)  and
  T2K (490~km/GeV).
  Bottom: Isocontours of P$_{\mu e}$ as a function of the two CP
  phases. The solid line corresponds to the standard value $P_{\mu
    e}^{3\times3}$ with $\delta_{CP} = 3\pi/2$ while colored regions
  denote the corresponding 10 and 20\% deviations, as indicated.}
\end{figure}

Notice that Eq.~(\ref{eq:Pmue}) represents in closed form the neutrino
transition probability \textit{in vacuo} in the presence of non--unitarity.
This expression bears some formal similarity to the Kuo--Pantaleone
formula~\cite{kuo:1989qe}.
The last term  in Eq.~(\ref{eq:Pmue}) is a small ``zero--distance'' effect 
characterizing the effective nonorthonormality
of the flavour neutrino states~\cite{Valle:1987gv}.
The corrections to the standard three--neutrino form are expected to be
small, howeve,r they involve a new CP phase, contained in the
interference term $P^I_{\mu e}$, so far unrestricted. Its explicit
form \textit{in vacuo} is given by
\begin{widetext}
\begin{equation}
P^{I}_{\mu e}  = 
-2 \sin2\theta_{13} \sin\theta_{23} \sin\Delta_{31}
   \sin\left(\Delta_{31} + \delta_{CP} +\phi \right)
   -   \cos\theta_{13} \cos\theta_{23} 
  \sin2\theta_{12}\sin2\Delta_{21}\sin\phi ,
\label{eq:PmueI}
\end{equation}
\end{widetext}
where we have set $\Delta_{ij} \equiv \frac{\Delta{
    m^2_{ij}L}}{4E_\nu}$. The CP violation phase--invariant parameter
 $\delta_{CP} =- (\phi_{12}-\phi_{13}+\phi_{23})$
 denotes the ``standard'' CP phase, while the CP violation phase
 associated with ``new physics'' is given as
 $\phi=\phi_{12}-\text{Arg}(\alpha_{21})$~\footnote{The $\phi_{ij}$
   are the phases associated to each complex rotation in the symmetric
   parametrization~\cite{Schechter:1980gr}.}.  The presence of this
 extra phase will lead to a degeneracy in the conversion probability.

 Notice that, for values of $L/E$ relevant for current and future long
 baseline neutrino experiments, the dependence of the appearance
 probability on the CP phases will be mainly determined by the
 interplay between two terms, one coming from the standard $P_{\mu
   e}^{3\times3}$, and the other one from the interference term
 $P_{\mu e}^I$, namely :
\footnotesize
\begin{eqnarray}
2 \alpha_{11}^2&\alpha_{22}^2& \sin\theta_{13}\sin\theta_{23}\sin\Delta_{31}
\sin 2 \Delta_{21}\times  \nonumber \\
& &\big[\sin2\theta_{12}\cos\theta_{23}\cos(\Delta_{31}+\delta_{CP}) \\
&-& 2 \frac{\cos\theta_{13}}{\alpha_{22}}\frac{|\alpha_{21}|}{\sin{2}\Delta_{21}}
\sin(\Delta_{31}+\delta_{CP}+\phi)\big] .  \nonumber 
\end{eqnarray}
\normalsize
By examining the brackets, one sees that, as expected, for vanishing
$\alpha_{21}$, we recover just the standard appearance probability,
while a relatively large $\alpha_{21}$ value clearly leads to a
degeneracy between $\delta_{CP}$ and $\phi$. 
 
This fact is illustrated in Fig.~\ref{fig:pemu-deg}, where we show the
conversion probability as a function of $L/E$ for different values of
the CP phases (top panel). One finds that, for a given $L/E$, the same conversion
probability can be obtained for several CP phase combinations.  Values
of $L/E$ for T2K, NO$\nu$A and DUNE are indicated with vertical lines
for illustration.  For the non--unitarity parameters we have
considered $\alpha_{11}^2 = \alpha_{22}^2 = 0.999$, and
$|\alpha_{21}|Ê= 2.5\times 10^{-2}$, consistent with the current
bounds obtained in \cite{Escrihuela:2015wra}.  All over the paper, the
neutrino oscillation parameters have been taken to their best fit
value obtained in Ref.~\cite{Forero:2014bxa}, with $\theta_{23}$ in the second
octant. Normal mass hierarchy has been assumed.
These degeneracies are further illustrated at the bottom panel of
Fig.~\ref{fig:pemu-deg} which shows the CP iso--contours that lead to
the same probability to within 10\% and 20\%, given a true value of
the standard three--neutrino probability with $\delta_{CP}=3\pi/2$, as
indicated by the current best fit point~\cite{Forero:2014bxa}.
In this figure we have fixed $L/E=500$~km/GeV which lies very close to the
value characterizing the T2K experiment. 

\section{Coping with the new ambiguity }

\begin{figure}[!b]
\centering
\includegraphics[width=0.4\textwidth]{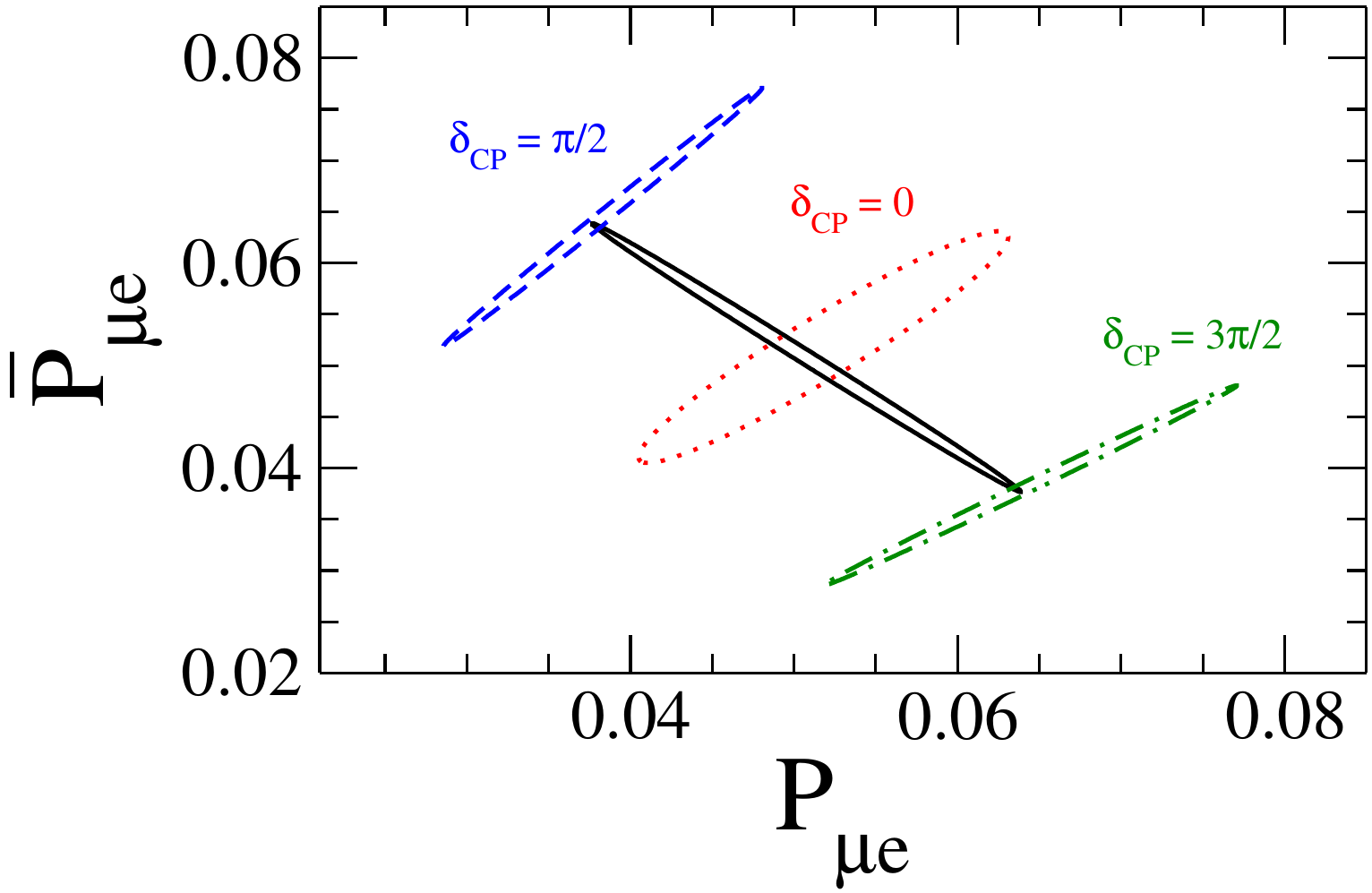}
\includegraphics[width=0.4\textwidth]{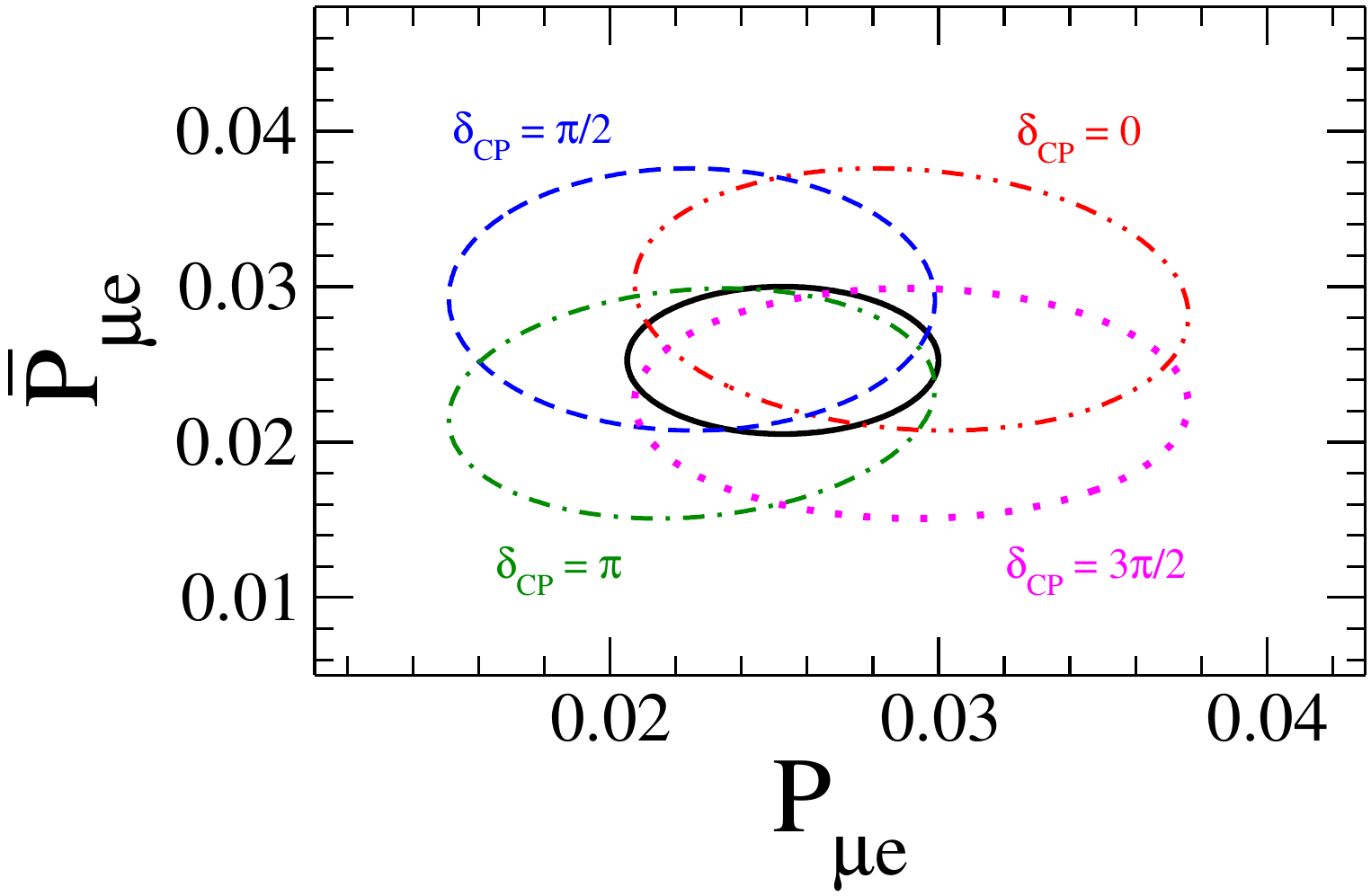}
\caption{Bi-probability plots for two different
  choices of $L/E$. The standard CP phase $\delta_{CP}$ is 
  fixed for each ellipse (except for the standard one denoted in black, where it varies freely),
   while the new phase $\phi$ is allowed 
  to vary from 0 to 2$\pi$.
  The upper panel, with $L/E=490$~km, corresponds to T2K while the
  bottom panel, with $L/E=250$~km, has been chosen for comparison.
  \label{fig:deg-T2K} }
\end{figure}

In Fig.~\ref{fig:pemu-deg} we saw how the new degeneracy associated with
non--unitarity leads to ambiguities in $P_{\mu e}$.
The comparison between the neutrino and the antineutrino channels
could provide a way to disentangle the CP phase $\delta_{CP}$ from the
new ``seesaw'' phase $\phi$ coming from non--unitarity.
Indeed, in the unitary case, the knowledge of a point ($P_{\mu e}$,
$\overline{P}_{\mu e}$) in the bi--probability plot will determine the
``standard'' CP phase up to the trigonometric $\delta_{CP} \to
\pi-\delta_{CP}$ ambiguity.

  In order to check whether this also holds true in the presence of
  non--unitarity we consider the bi--probability plots in
  Fig.~\ref{fig:deg-T2K}.
  The upper panel shows that, for values of $L/E$ close to 500 km/GeV,
  the combination of neutrino and antineutrino measurements removes
  the degeneracies between the CP phases present in each channel
  separately.  In fact, this can be understood from a detailed
  analysis of the CP--dependent terms in $P_{\mu e}$ as given by
  Eq.~(\ref{eq:Pmue}). One finds that, for $L/E$ = 500 km/GeV, some of
  these terms cancel exactly. The degeneracies in the phases
  $\delta_{CP}$ and $\phi$ due to the remaining terms, present in both
  the neutrino and antineutrino channels separately, disappear once
  the two channels are combined.  Fortunately, neutrino long-baseline
  experiments are usually tuned to the ratio $L/E=500$~km/GeV, where
  the oscillation maximum is located.
  
    However, for $L/E$ values far from 500 km/GeV, the interplay
    between the different CP--dependent terms in $P_{\mu e}$ is rather
    involved. As a result, the phase degeneracies present in the
    neutrino channel may persist even after the combined two--channel
    analysis including antineutrino observations.
  Indeed, as can be seen in the bottom panel of
  Fig.~\ref{fig:deg-T2K}, the ambiguities in general remain even with
  the combined measurements of the appearance probabilities in
  neutrino ($P_{\mu e}$) and antineutrino channel ($\overline{P}_{\mu
    e}$). 
  Therefore, the conventional strategy will not in general be enough
  to ensure an unambiguous determination of the ``standard'' CP
  phase in the present case.

\begin{figure} \centering
\includegraphics[width=0.35\textwidth]{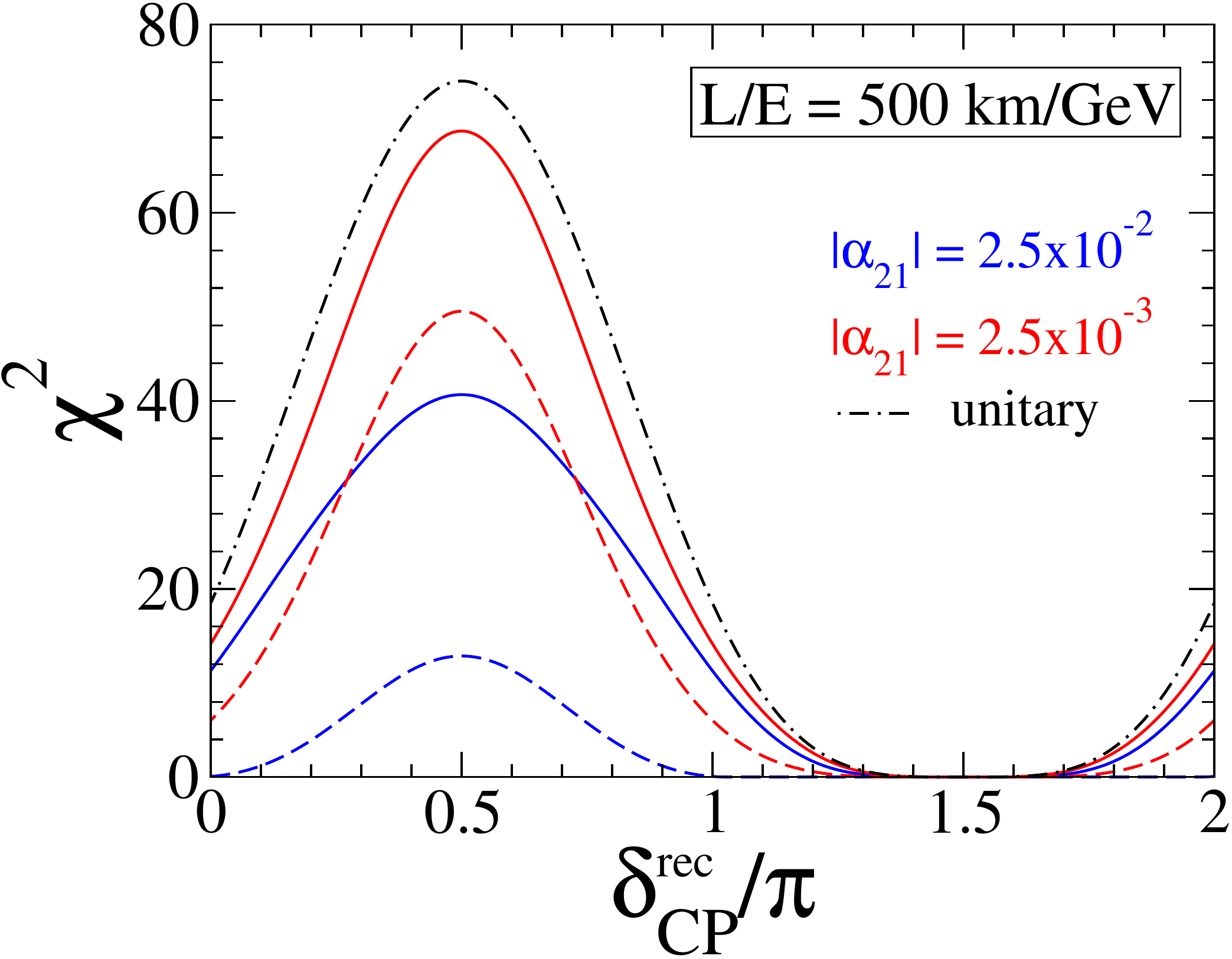}
 \includegraphics[width=0.35\textwidth]{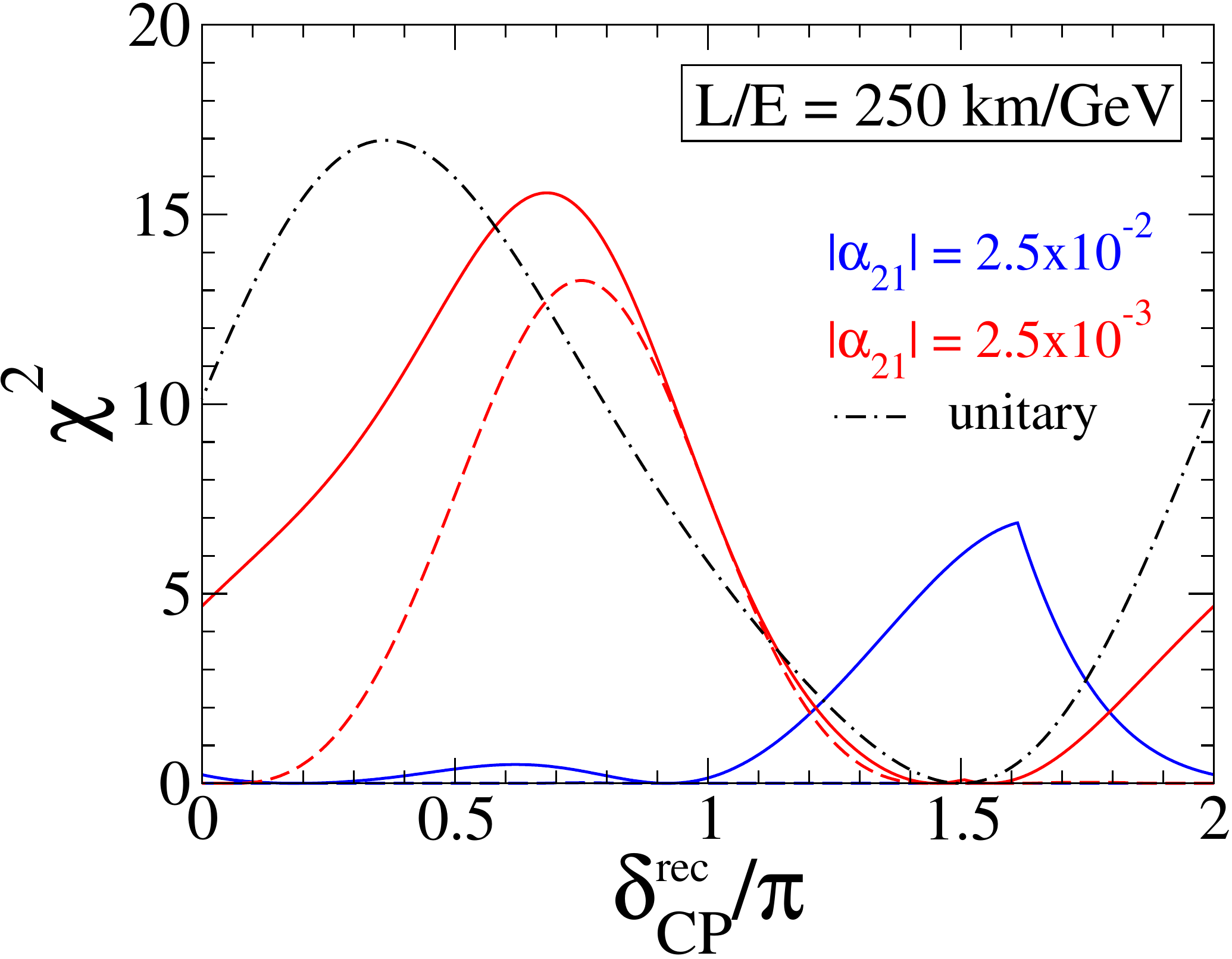}
 \caption {$\delta_\mathrm{CP}$ determination in the presence of
   non--unitarity for two different values of $L/E$ and
   $|\alpha_{21}|$.  The true value of $\delta_\mathrm{CP}$ has been
   taken equal to 3$\pi$/2. Dashed (solid) lines correspond to the
   reconstructed value from the neutrino (combined neutrino +
   antineutrino) appearance probability. The dot--dashed line
     shows the reconstruction of the standard CP phase in the unitary
     case.\label{fig:recons}}
    \end{figure}

  Likewise, one can obtain a quantitative measure of the
  reconstruction sensitivity of the ``standard'' phase $\delta_{CP}$
  in the presence of non--unitarity, as shown in
  Fig.~\ref{fig:recons}~\footnote{ These results have been obtained by
    fitting the neutrino oscillation probability, assumed to be
    measured with a 10\% uncertainty.}
    One finds that by combining the two channels the reconstruction is
    very much improved and is close to that obtained in the standard
    unitary case, just a bit worse due to the presence of the extra
    degree of freedom $\phi$. This holds for $L/E$=500 km/GeV or
    close.  In contrast, for $L/E$ values far from the above, say 250
    km/GeV, the reconstruction sensitivity is lost completely.
    Indeed, with the neutrino channel alone one has no sensitivity at
    all, with the corresponding dashed blue line being hardly visible,
    overlapping the horizontal axis. Combining neutrino and
    antineutrino channels does not solve the situation, as a local
    $\chi^2$ maximum appears at $\delta_{CP}$ = $3\pi/2$, the true
    simulated value.
    One also finds how for $\alpha_{21} \to 0$ the standard case is
    recovered, lifting the new degeneracy relatively well for $L/E$
    500 km/GeV and $\alpha_{21} < 2.5 \times 10^{-3}$ (top
    panel). Unfortunately, however, stringent direct limits on
    $\alpha_{21}$ are inexistent. There are only indirect restrictions
    from charged \lfv processes, difficult to quantify in a
    model--independent way. For a recent discussion in the context of
    seesaw models see Ref.~\cite{Forero:2011pc}{ \footnote
           In that paper it was shown that values of $\alpha_{21}$
          up to $3\times 10^{-3}$ are in agreement with constraints
          from LFV searches at 90\% CL.  However, those bounds hold
          within a restrictive ``minimal ansatz''. Here we prefer to
          be conservative and apply only the truly model--independent
          bounds on $\alpha_{21}$ derived in
          Ref.~\cite{Escrihuela:2015wra}.}

We must stress that, for simplicity, we have our study to restricted
to neutrino oscillations \textit{in vacuo}.
This is reasonable because we are focussing on degeneracies associated
with intrinsic CP violation. In this sense it is relevant to
investigate whether the vacuum probabilities provide a robust
signature of CP violation. Although the inclusion of matter effects
will be necessary for realistic predictions for very long baseline
experiments such as DUNE~\cite{Acciarri:2015uup}, it is expected to
modify but not destroy the existence of the new degeneracies noted here.

\section{Conclusions}
We have argued, on the basis of the seesaw mechanism, that the lepton
mixing matrix describing neutrino oscillations is likely to be
non--unitary. We have focussed on the description of neutrino
oscillations in the non--unitary case, in particular on the effects of
the extra CP phase present in the oscillation probabilities.
We have identified degeneracies in the appearance probability $P_{\mu
  e}$ for different combinations of the ``standard'' three--neutrino phase
$\delta_{CP}$ and the ``new'' CP phase $\phi$ associated with
 the new parameters describing non--unitarity.
 These ambiguities are beyond the conventional ones, having no
 analogue within the standard unitary three--neutrino oscillation
 benchmark.
 We have discussed the resulting ambiguities in oscillation
 probabilities for various $L/E$ and non--unitarity parameter choices.
We have outlined the simplest strategies to help coping with the
presence of these new degeneracies.
The standard strategy of determining $\delta_{CP}$ from the
combination of neutrino and antineutrino observations, that holds in
the unitary case, turns out to be insufficient in removing the degeneracies
between two CP--phases $\delta_{CP}$ and $\phi$ for values of $L/E$
far from the ``magic'' value of 500 km/GeV.
In short, we showed how ``generic'' neutrino oscillation measurements
are not individually robust with respect to unitarity violation
effects expected within a class of seesaw schemes.
New strategies and/or combined studies using data from different
experiments may be necessary in order to ensure unambiguous CP
measurements.   Such efforts offer a valuable window for
complementary tests of \lfc and weak universality.
Before closing let us also mention that CP ambiguities
will also arise within generic non--standard interaction schemes not
directly related to a seesaw mechanism as the origin of neutrino
mass. Likewise, dedicated studies, analogous to those
in~\cite{Forero:2016cmb,deGouvea:2015ndi,Coloma:2015kiu,Masud:2016bvp}
will be required here in order to cover each experimental setup.

\black
\section*{Acknowledgements}

Work supported by Spanish grants FPA2014-58183-P, Multidark
CSD2009-00064, SEV-2014-0398 (MINECO), PROMETEOII/2014/084
(Generalitat Valenciana), and the CONACyT grant 166639.
M.~T. is supported by a Ram\'{o}n y Cajal contract (MINECO).

\bibliographystyle{unsrt}

\end{document}